%% file: main.tex
\PassOptionsToPackage{table}{xcolor}
\documentclass[10pt,conference]{IEEEtran}
\usepackage{cite}
\usepackage{amsmath,amssymb,amsfonts}
\usepackage{graphicx}
\usepackage{textcomp}
\usepackage{xcolor}
\usepackage{booktabs}
\input{includes}

\hypersetup{hidelinks}

\newsavebox{\persesbox}
\newsavebox{\scythebox}

\begin{document}
\sloppy

\title{Scythe: Semantic-Aware Rewriting for Fast,\\Effective Program Reduction}

\author{%
\IEEEauthorblockN{Ioanna Moraiti\textsuperscript{1},
Charalambos Mitropoulos\textsuperscript{2},
Artem Anchikov\textsuperscript{3},
Sotirios Ioannidis\textsuperscript{2}\\
Dimitris Mitropoulos\textsuperscript{4,5},
Thodoris Sotiropoulos\textsuperscript{1}}
\IEEEauthorblockA{\textsuperscript{1}Unaffiliated,
\textsuperscript{2}Technical University of Crete,
\textsuperscript{3}ETH Zurich,\\
\textsuperscript{4}University of Athens,
\textsuperscript{5}National Infrastructures for Research and Technology}
\IEEEauthorblockA{\{ioannamoraiti@outlook.com, theosotr@windowslive.com\},
\{cmitropoulos, sioannidis\}@tuc.gr,
aanchikov@ethz.ch,
dimitro@ba.uoa.gr}
}

\maketitle

\begin{abstract}
Automated program reduction shrinks
a program into a smaller one
that preserves a property of interest.
Language-agnostic grammar-based reducers are easy to
retarget across languages,
but because they reason only about syntax,
they cannot perform the coupled edits
needed to remove whole declarations,
e.g., a method cannot be deleted while it is still called. 
As a result,
they waste oracle calls on edits
that cannot succeed
and often leave irrelevant declarations behind.
We present~\tool,
a~\emph{fast},
\emph{semantic-aware}~\emph{pre-processing} step
that runs in front of any grammar-based reducer without modifying it.
\tool encodes the program as a typed declaration dependency graph
and performs a set of semantic-aware operators
that decouple declarations from their use sites,
making them removable while keeping the program well-typed.
Fine-grained token-level minimization is then left to
the downstream reducer.
On 73 benchmarks across C, Java, and Solidity,
combining~\tool with~\perses never yields a larger result
than~\perses alone,
and a strictly smaller one
on~\empirical{70} of~\empirical{73} cases.
On average,
the combination produces programs~\empirical{55.6\%} smaller
(up to~\empirical{95.5\%})
and is~$\empirical{2.3}\times$ faster
(up to~$\empirical{7.4}\times$).
\tool also boosts~\latra,
a reducer that,
unlike the purely syntactic~\perses,
also applies language-specific transformations:
the combination produces programs~\empirical{10.6\%} smaller
(up to~\empirical{61.8\%})
and is~$\empirical{2.4}\times$ faster on average.
Video: \url{https://youtu.be/uEq_YYU0zZw}
\end{abstract}

\begin{IEEEkeywords}
Program reduction, delta debugging, compiler testing,
semantics
\end{IEEEkeywords}

\section{Introduction}
\label{sec:intro}

Automated program reduction is integral to
debugging and bug deduplication in
software systems that take programs as input,
such as compilers,
interpreters,
and static analyzers.
Given a program $P$ that exhibits a property of interest,
such as a compiler crash,
a miscompilation,
or a static-analysis report,
a reducer produces a smaller program $P'$
that still exhibits it,
querying an~\emph{oracle}
(a script that checks the property)
on each candidate~\cite{creduce}.
Existing reducers fall into two families.
Language-specific reducers,
such as C-Reduce~\cite{creduce},
apply transformations tailored to one language (e.g., C).
These transformations make reduction effective,
but the tools are hard to retarget to other languages.
Language-agnostic,
grammar-based reducers~\cite{perses,vulcan}
instead delete syntax
while guaranteeing that the result still parses,
so they can be retargeted to a new language
simply by supplying its grammar.
However,
this generality comes at a cost.
Language-agnostic reducers can be slow,
even on small programs,
and often leave behind code
that is irrelevant to the property of interest.

\point{Example}
Consider a Java program that crashes the Java compiler
({\tt javac}, JDK 8.0.32);
it spans 300 lines with 16 top-level types,
55 methods,
and heavy use of inheritance,
generics,
and lambdas.
Figure~\ref{fig:motivating-perses} shows
the version produced by~\perses~\cite{perses},
a state-of-the-art language-agnostic reducer,
after roughly five minutes.
Several remaining declarations
(e.g., class {\tt Base}, field {\tt active}, method {\tt label})
are irrelevant to the bug,
yet~\perses cannot remove them
because they are still referenced elsewhere.

\point{The problem}
Programs targeted by reduction tools contain
many declarations unrelated to the property of interest
(here, most of the 16 types and 55 methods).
The goal is to
\textit{identify and remove these irrelevant declarations,
ideally eliminating large blocks of code
quickly with few oracle queries}.
However,
doing so requires breaking the
dependencies that tie each declaration to its use sites,
edits that a language-agnostic reducer cannot perform.

\point{Our tool}
We bridge the gap between language-agnostic
and language-specific reducers with~\tool,
a~\textit{fast},
\textit{semantic-aware}~\textit{pre-processing} step
that runs before a language-agnostic reducer.
It targets developers and testers
of compilers, interpreters, and static analyzers,
e.g., when reducing fuzzer-found bugs.
\tool removes whole classes,
methods,
and fields,
applying a set of~\textit{general,
semantic-aware decoupling operators}
that rewrite a declaration's use sites.
The declaration then becomes removable
while the program remains well-typed.
\tool supports C,
Java,
and Solidity,
and is easily extensible to new languages.
On our running example,
the~\tool $+$~\perses pipeline produces
the smaller program of Figure~\ref{fig:motivating-scythe} in
about two minutes
($\empirical{2.1}\times$ speedup),
with~\emph{43\%} fewer tokens than~\perses alone.
Across~\empirical{73} benchmarks,
\tool never makes the output of~\perses larger,
at least halves it in~\empirical{50} cases,
and speeds up reduction by up to~$\empirical{7.4}\times$.
On Java,
\tool alone,
in under a minute on average,
already beats~\perses on~\empirical{23} of~\empirical{34} benchmarks.
Paired with~\latra~\cite{latra},
\tool yields programs~\empirical{10.6\%} smaller
(up to~\empirical{61.8\%})
and reduction~$\empirical{2.4}\times$ faster on average,
all~\emph{without} modifying
the underlying reducers.

\section{Challenges}
\label{sec:background}

We identify two challenges that any reducer must overcome
to remove irrelevant declarations
(e.g., class {\tt Base}, Figure~\ref{fig:motivating-perses}).

\begin{figure}[t]
\centering
\begin{lrbox}{\persesbox}
\begin{minipage}{\linewidth}
\begin{lstlisting}[style=javapanel]
class Main {
  Widget active;
  static String label(short code) { return "tag"; }
  final boolean READY =
    new Widget("first").check(active, () -> {
      String text = "x"; return text; })
    || active.check(new Widget(Main.label((short) 54)), null);
}
interface Supplier<R> { R get(); }
class Base { Base(Boolean flag, boolean extra) { } }
class Widget extends Base {
  Widget(String tag) { super(false, true); }
  Boolean check(Widget other, Supplier task) { return true; }
}
\end{lstlisting}
\end{minipage}
\end{lrbox}
\begin{lrbox}{\scythebox}
\begin{minipage}{\linewidth}
\begin{lstlisting}[style=javapanel]
class Main {
  final boolean READY =
    ((__A) null).check(null, () -> {
      String text = "x"; return text;
    })
    || ((__A) null).check(null, null);
}
interface Supplier<R> { R get(); }
class __A {
  Boolean check(__A other, Supplier task) { return true; }
}
\end{lstlisting}
\end{minipage}
\end{lrbox}
\subfloat[]{\usebox{\persesbox}\label{fig:motivating-perses}}

\subfloat[]{\usebox{\scythebox}\label{fig:motivating-scythe}}
\caption{
An example of a Java program that crashes JDK 8.
(a) The minimized program produced by~\perses,
a language-agnostic reducer;
(b) The minimized program produced by~\perses after 
the code is pre-processed by~\tool.
}
\label{fig:motivating}
\end{figure}

\point{Challenge 1:
program dependencies that block removal}
Pruning a declaration usually makes
the oracle reject the program,
even when the declaration is irrelevant to the bug,
because it is referenced at multiple sites.
For example,
pruning class {\tt Base} alone makes the crash disappear,
since {\tt Base} still appears after {\tt extends} (line 11)
and its constructor is called (line 12).
A reducer must therefore identify
every~\emph{dependency}
between a declaration and its uses,
and different uses demand different edits.
Removing the method {\tt label} requires
dropping or replacing its call.
Removing a class used as the type of a field,
such as {\tt Widget} for {\tt active} (line 2),
may require removing the field
and, in turn, every use of that field.

\point{Challenge 2: Syntax-guided search is prohibitively expensive}
Existing reducers employ sophisticated search to find
further reduction opportunities.
\vulcan~\cite{vulcan} guarantees tree-minimality,
a stronger property than 1-minimality,
through exhaustive,
\emph{syntax-guided} search
that can discover that removing {\tt Base} also requires
removing its {\tt super} call and the inheritance.
However,
this is very costly
and still misses opportunities~\cite{latra}:
\vulcan cannot reduce our example
even after six hours.

\section{\tool}
\label{sec:approach}

\begin{figure}[t]
\centering
\includegraphics[width=0.7\linewidth]{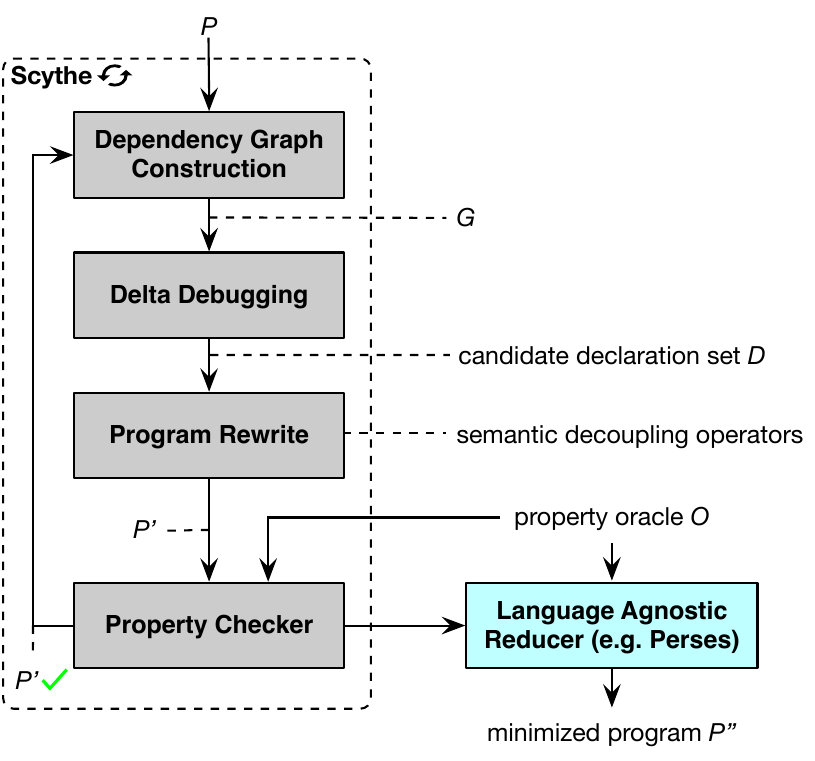}
\caption{
The high-level architecture of~\tool.
}
\label{fig:arch}
\end{figure}

\point{Overview}
Figure~\ref{fig:arch} shows
the high-level architecture of~\tool,
which runs as a fast pre-processing step
\emph{before} the actual language-agnostic reducer.
Its goal is~\emph{not} to produce
a fully minimized program
or to remove every redundant token,
but to quickly remove bulk code
(\textit{Challenge 2})
and whole declarations
that the downstream reducer would struggle to reason about
(\textit{Challenge 1}).
\tool takes a source file $P$
and a property oracle $O$,
and outputs a program $P'$
for the downstream reducer.
The two may alternate until the program stops shrinking,
as each can expose new opportunities for the other.
Internally,
\tool iterates four steps until a fixpoint.

\begin{figure}[t]
\centering
\includegraphics[width=0.7\linewidth]{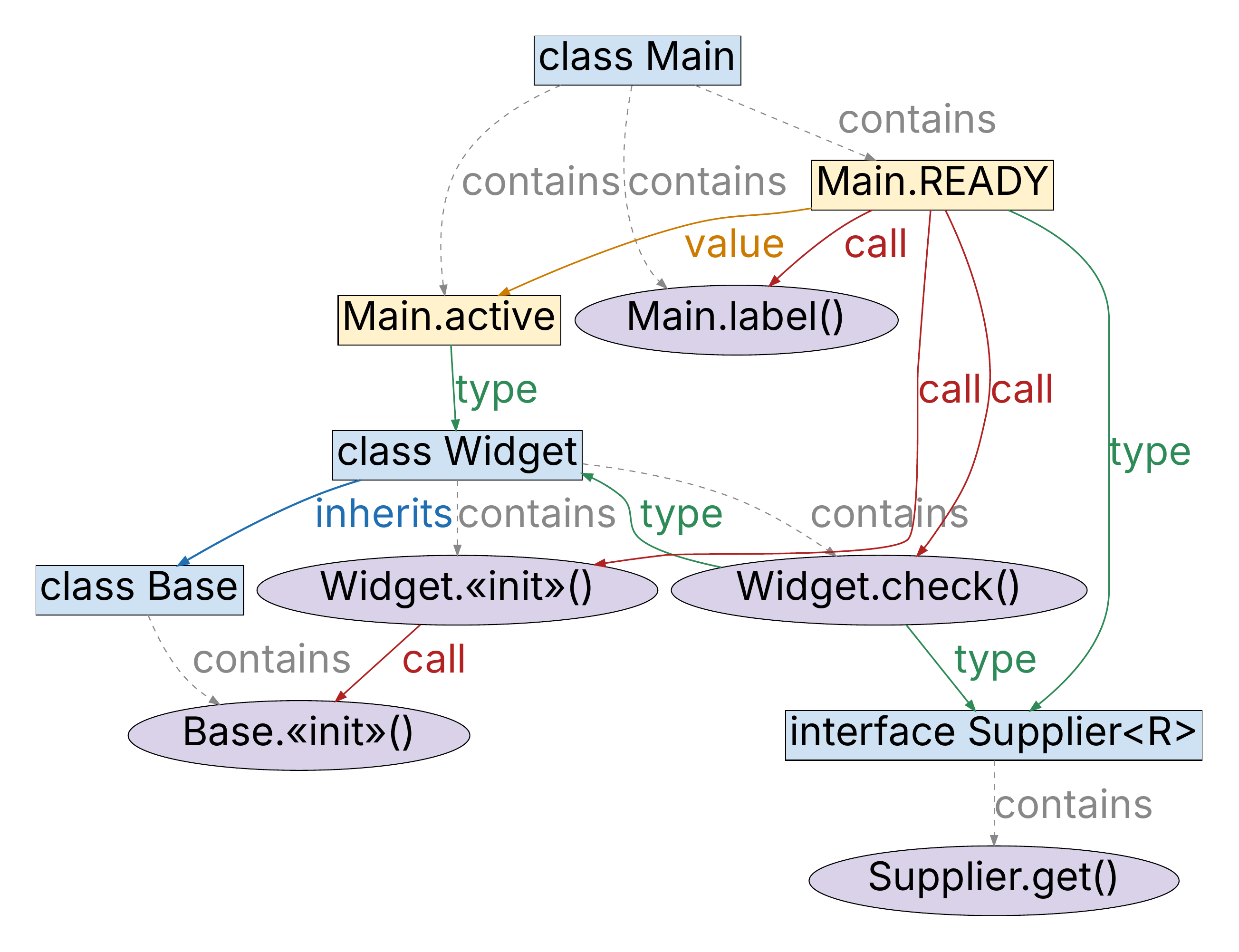}
\caption{
The dependency graph of the program of Figure~\ref{fig:motivating-perses}.
}
\label{fig:graph}
\end{figure}

\point{Step 1: dependency graph construction}
\tool parses the input program $P$
into a graph
that we call the~\textit{typed dependency graph}.
Each node represents a declaration,
such as a function,
field,
or class.
Each edge records how one declaration uses another.
The edge labels let~\tool pick the right way to
break each dependency before removing a declaration
(see~\textit{Step 3}).
The graph (Figure~\ref{fig:graph}) captures four kinds of relationships:
(1)~\textit{containment}
({\tt Main} contains {\tt Main.active}),
(2)~\textit{inheritance}
({\tt Widget} inherits from {\tt Base}),
(3)~\textit{type}
({\tt Main.active} has type {\tt Widget}),
and (4)~\textit{value/call}
({\tt Main.READY} reads {\tt Main.active}).

\point{Step 2: delta debugging}
\tool applies delta debugging~\cite{ddmin}
over the nodes of $G$,
i.e., at the level of declarations.
Rather than a single reduction list over all nodes,
\tool\ schedules a sequence of passes,
each on a single declaration type,
ordered from coarse to fine-grained
(e.g., classes, then methods, then fields).
Deleting a class discards all its nested code
in one oracle call.
Each pass yields a candidate set $D \subseteq V(G)$,
tested against the oracle (Step 4).
The process guarantees 1-minimality
at the level of declarations.

\begin{table}[t]
  \centering
  \caption{
      \tool's semantic decoupling operators.
  }
  \label{tab:operators}
  \setlength{\tabcolsep}{5pt}
  \resizebox{0.9\linewidth}{!}{
  \begin{tabular}{@{}lll@{}}
    \toprule
    \textbf{Declaration $d$} & \textbf{Use Site $u$} & \\
    \textbf{(Kind)} & \textbf{(Kind)} & \textbf{Operator} \\
    \midrule
    \multirow{3}{*}{Class}
      & {\tt inherits} & $\textsc{Remove}(u)$ \\[2pt]
      & {\tt type}   &  $\textsc{AddClass}(u, \texttt{\_\_A});$ \\
      &              &  $\textsc{Replace}(u,\ \texttt{\_\_A})$ \\
    \midrule
    \multirow{2}{*}{Function}
      & {\tt call} (result used)    & $\textsc{Replace}\bigl(u,\ \textsc{Default}(\textit{ret}(d))\bigr)$ \\[2pt]
      & {\tt call} (unused result) & $\textsc{Remove}(u)$ \\
    \midrule
      Any & {\tt value}    & $\textsc{Replace}\bigl(u,\ \textsc{Default}(\textit{type}(d))\bigr)$ \\
    \midrule
      Any      & {\tt contains} & $\textsc{Remove}(d)$ \\
    \bottomrule
  \end{tabular}}
\end{table}

\point{Step 3: program rewrite}
For each declaration in the candidate set $D$,
\tool removes the declaration node
and applies a set of~\textit{semantic-aware decoupling operators}
that detach it from its use sites.
To decide which operators to apply,
\tool examines the incoming edges of the node
(its uses),
together with those of the declarations it contains.
The kind of reference selects the operator for each use
(Table~\ref{tab:operators}).

For example,
to remove class {\tt Base}
(Figure~\ref{fig:motivating-perses}, line 10),
\tool applies {\sc Remove} to its {\tt inherits} edge from {\tt Widget},
dropping the {\tt extends} clause,
and to the call of its constructor
from {\tt Widget}'s constructor,
whose result is unused.
With both edges gone,
{\tt Base} and its constructor are deleted
(Figure~\ref{fig:motivating-scythe}).
Similarly,
a type use of a removed class is rewritten to
a placeholder class {\tt \_\_A}
({\sc AddClass}; {\sc Replace}).
A value use,
or a call whose result is used as an expression,
is replaced by the default value of the corresponding type
($\textsc{Default}(\textit{type}(d))$).
This is why the references to {\tt Widget} and
the field {\tt active} collapse to
\texttt{((\_\_A) null)} in Figure~\ref{fig:motivating-scythe}
(lines 3, 6).

\point{Step 4: property checking}
If the rewritten $P'$ still satisfies the oracle $O$,
the change is committed (feedback arrow in Figure~\ref{fig:arch});
otherwise,
it is rolled back.
Faulty rewrites thus never reach the output,
assuming,
like any reducer,
a deterministic oracle.

\subsection{Implementation, Usage, and Availability}
\tool is implemented in $\sim$\num{4000} lines of Python.
Parsing is delegated to
{\tt tree-sitter}~\cite{tree-sitter};
adding a new language only requires
mapping {\tt tree-sitter} node types to
\tool's declaration and edge kinds
and instantiating the operators of Table~\ref{tab:operators}.
\tool relies on the {\tt picire}~\cite{picire}
library for delta debugging,
but could be easily configured to use any list reduction algorithm
(e.g., probabilistic delta debugging~\cite{probdd}).
Users pass the input program,
the oracle,
and, optionally,
a wrapper for the downstream reducer:
\begin{lstlisting}[language={},numbers=none,xleftmargin=0pt,basicstyle=\footnotesize\ttfamily]
scythe --language java --source-file A.java \
  --script oracle.sh --post-processor perses.sh
\end{lstlisting}
\tool is open source (GPL-3.0) at
\url{https://github.com/itiarom/scythe},
with a Docker image and usage instructions;
the artifact is at
\url{https://doi.org/10.5281/zenodo.23589196}.

\section{Preliminary Evaluation}
\label{sec:evaluation}

We evaluate whether~\tool improves
the (\textbf{RQ1}) quality
and (\textbf{RQ2})
efficiency of language-agnostic reducers.

\point{Benchmarks}
No public dataset spans C, Java, and Solidity,
so we collect 73 programs from three sources.
\textit{C (20 benchmarks)}:
the \perses benchmarks~\cite{perses},
Csmith~\cite{csmith}-generated programs
that trigger crashes and miscompilations
in GCC and LLVM,
widely used in prior work~\cite{perses,latra,trec}.
\textit{Java (34 benchmarks)}:
we generate programs with \textsc{Hephaestus}~\cite{hephaestus}
against OpenJDK:
20 programs that {\tt javac} (JDK 11) rejects
with an unexpected compile-time error
and 14 that crash {\tt javac} (JDK 8).
\textit{Solidity (19 benchmarks)}:
smart contracts from
the dataset of Mitropoulos et al.~\cite{MKMISM23},
each triggering a different warning or error
in the Slither static analyzer~\cite{slither}.

\point{Reducers}
We pair~\tool with two state-of-the-art reducers:
the grammar-based~\perses~\cite{perses}
and~\latra~\cite{latra},
which adds language-specific transformation templates
and performs comparably to C-Reduce.
We run~\latra only on C,
as it does not support Java or Solidity.
We exclude C-Reduce~\cite{creduce}:
it is C-specific,
whereas~\tool is designed to
\textit{boost language-agnostic reducers,
so the two are complementary rather than competing}.
For each benchmark and reducer~R,
we run three pipelines:
(1)~\tool alone (\textbf{S}),
(2)~R alone (\textbf{P} or~\textbf{L}),
and (3)~\tool followed by~R (\textbf{S+P} or~\textbf{S+L}).

\begin{table}[t]
\centering
\caption{
Sample results: size in tokens (\textit{Tok}), time in seconds (\textit{s}),
and bug type (\textit{Prop}): \textit{U} (unexpected error),
\textit{CR} (crash), \textit{M} (miscompilation).
}
\label{tab:results}
  \setlength{\tabcolsep}{3.5pt}
  \resizebox{0.8\linewidth}{!}{
  \begin{tabular}{@{}lcrrrrrrr@{}}
  \toprule
   & & & \multicolumn{2}{c}{\textsc{S}} & \multicolumn{2}{c}{\textsc{P}} & \multicolumn{2}{c}{\gp} \\
  \cmidrule(lr){4-5}\cmidrule(lr){6-7}\cmidrule(lr){8-9}
  Benchmark & Prop & Init & Tok & $s$ & Tok & $s$ & Tok & $s$ \\
  \midrule
  \multicolumn{9}{@{}l}{\textit{C}} \\
   \texttt{clang-31259} & M & \num{48079} & \num{2101} & \num{744} & \num{518} & \num{4111} & \bcell{\num{240}} & \bcell{\num{3546}} \\
   \texttt{gcc-64990} & M & \num{147002} & \num{1861} & \num{1807} & \num{435} & \num{4624} & \bcell{\num{104}} & \bcell{\num{3687}} \\
   \texttt{gcc-61917} & CR & \num{84473} & \num{2471} & \num{2515} & \num{579} & \num{5956} & \bcell{\num{125}} & \bcell{\num{4177}} \\
   \texttt{gcc-59903} & CR & \num{56951} & \num{2635} & \num{1839} & \num{722} & \num{5171} & \bcell{\num{302}} & \bcell{\num{4723}} \\
  \midrule
  \multicolumn{9}{@{}l}{\textit{Java}} \\
   \texttt{jdk11-10} & U & \num{5298} & \num{83} & \num{98} & \num{657} & \num{731} & \bcell{\num{31}} & \bcell{\num{124}} \\
   \texttt{jdk11-7} & U & \num{3158} & \num{80} & \num{55} & \num{382} & \num{343} & \bcell{\num{31}} & \bcell{\num{81}} \\
   \texttt{jdk8-24} & CR & \num{4691} & \num{131} & \num{73} & \num{608} & \num{796} & \bcell{\num{74}} & \bcell{\num{121}} \\
   \texttt{jdk8-34} & CR & \num{3377} & \num{143} & \num{44} & \num{591} & \num{649} & \bcell{\num{95}} & \bcell{\num{116}} \\
  \midrule
  \multicolumn{9}{@{}l}{\textit{Solidity}} \\
   \texttt{smart17} & U & \num{2570} & \num{69} & \num{65} & \num{93} & \num{298} & \bcell{\num{16}} & \bcell{\num{86}} \\
   \texttt{smart7} & U & \num{4103} & \num{143} & \num{60} & \num{92} & \num{325} & \bcell{\num{33}} & \bcell{\num{133}} \\
   \texttt{smart13} & U & \num{1500} & \num{55} & \num{48} & \num{84} & \num{161} & \bcell{\num{38}} & \bcell{\num{91}} \\
   \texttt{smart4} & U & \num{1288} & \num{100} & \num{22} & \num{64} & \num{122} & \bcell{\num{29}} & \bcell{\num{59}} \\
  \bottomrule
  \end{tabular}}
\end{table}

\begin{figure}[t]
\centering
\includegraphics[width=0.8\linewidth]{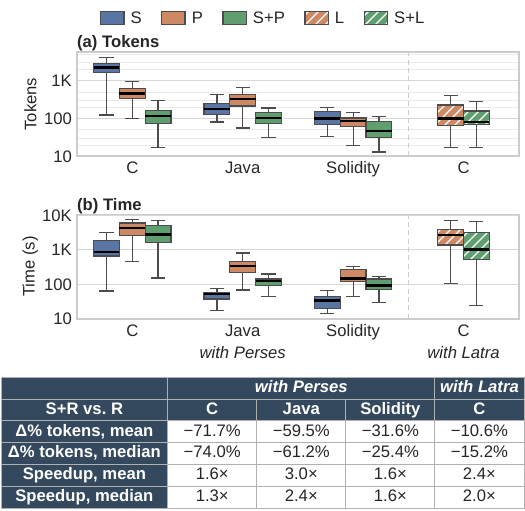}
\caption{Reduction size (a) and time (b) of \tool alone (S),
a reducer R alone (P: \perses; L: \latra, hatched),
and \tool followed by R (S$+$R).}
\label{fig:boxplots}
\end{figure}

\point{Results}
Table~\ref{tab:results} shows
four benchmarks per language
(all are in our artifact),
and Figure~\ref{fig:boxplots} shows
the distributions across all benchmarks.
\tool's own runtime is small
(\empirical{14} minutes,
50 seconds,
and 30 seconds on C,
Java,
and Solidity).
\textbf{S+P} substantially
reduces final program size by
\empirical{71.7\%},
\empirical{59.5\%},
and~\empirical{31.6\%} on average
for C,
Java,
and Solidity,
respectively.
\textbf{S+P} accelerates reduction by
$\empirical{1.6}\times$,
$\empirical{3}\times$,
and~$\empirical{1.6}\times$
on average for C,
Java,
and Solidity,
respectively.
On C,
this saves \empirical{23} minutes per benchmark on average
and up to \empirical{2.5} hours.
Overall,
\textbf{S+P} is never larger than~\textbf{P},
strictly smaller on~\empirical{70} of~\empirical{73} benchmarks,
and faster on~\empirical{67}.

\point{Beyond~\perses}
The benefits carry over to~\latra:
\textbf{S+L} is~$\empirical{2.4}\times$ faster than~\textbf{L} on average
(up to~$\empirical{7.3}\times$),
cutting total time
from~\empirical{19.0} to~\empirical{11.4} hours.
Although~\latra already applies
language-specific transformations,
\textbf{S+L} yields smaller programs than~\textbf{L}
in~\empirical{13} cases and equal ones in~\empirical{one}
(\empirical{15.2\%} smaller at the median,
up to~\empirical{61.8\%}).
In~\empirical{six} cases,
the output grows by~\empirical{3}--\empirical{86} tokens,
as \tool's default values (Table~\ref{tab:operators}),
e.g., \texttt{0xDEADBEEF} for an \texttt{int},
can hinder~\latra's transformations;
we plan to fix this in future work (Section~\ref{sec:conclusion}).

\section{Related Work}
\label{sec:related-work}

The closest line of work is grammar-guided reduction
over a syntactically valid parse tree.
\perses~\cite{perses} introduced the quotient
representation that keeps every candidate parseable,
\textsc{Pardis}~\cite{pardis} prioritises subtrees by
expected gain,
and type-batched reduction~\cite{tbpr}
amortises oracle cost across simultaneous queries. All
operate \emph{within} a single parse tree, whereas \tool runs one
step \emph{before} them, pruning whole declarations on a
graph abstraction, and composes with any of them.

The remaining reducers fall into two groups.
Delta-debugging engines
(\texttt{ddmin}~\cite{ddmin}, HDD~\cite{hdd},
\textsc{Picire}~\cite{picire},
\textsc{ProbDD}~\cite{probdd}) improve the
underlying search policy and are orthogonal to \tool: any
can replace its delta-debugging core. Language-specific reducers
(\textsc{C-Reduce}~\cite{creduce}, \textsc{GTR}~\cite{gtr},
\textsc{LPR}~\cite{lpr}) encode per-language
passes (hand-coded, learned, or LLM-suggested) and reach
smaller results, but work in only one language.
\tool complements both families, and even improves
\latra, which performs on par with C-Reduce.

\section{Conclusion and Future Work}
\label{sec:conclusion}

We presented~\tool,
a semantic-aware front-end
that makes language-agnostic reducers
both faster and more effective,
without modifying them.
Paired with~\perses,
\tool never produced a larger result
across~\empirical{73} benchmarks in three languages,
halved the output in~\empirical{50} of them,
and cut total reduction time
from~\empirical{31.2} to~\empirical{21.0} hours,
lowering the compute and energy cost of debugging.
Beyond this preliminary evaluation,
we plan to select operators (Table~\ref{tab:operators})
and default values context-sensitively,
to open further opportunities downstream
and avoid the~\latra regressions.

\bibliographystyle{IEEEtran}
\bibliography{main}

\end{document}

%% file: includes.tex
\usepackage{url}
\usepackage{listings}
\usepackage[linesnumbered,ruled,noend]{algorithm2e}
\usepackage[caption=false,font=footnotesize]{subfig}
\usepackage{multirow}
\usepackage{multicol}

\usepackage{soul}
\usepackage{caption}

\makeatletter
\IfFileExists{siunitx.sty}{\@tempswatrue}{\@tempswafalse}
\makeatother
\makeatletter
\if@tempswa
  \usepackage{siunitx}
\else
  \providecommand{\num}[1]{#1}
\fi
\makeatother

\SetKw{Continue}{continue}

\PassOptionsToPackage{hyphens}{url}\usepackage{hyperref}

\definecolor{dkred}{RGB}{87,10,10}
\definecolor{burntorange}{rgb}{0.8, 0.33, 0.0}
\definecolor{OrangeRed}{rgb}{1.0, 0.27, 0.0}
\definecolor{ForestGreen}{rgb}{0.0, 0.27, 0.13}

\definecolor{groovyblue}{HTML}{0000A0}
\definecolor{dkgreen}{HTML}{008000}
\definecolor{darkgray}{rgb}{.4,.4,.4}
\definecolor{bittersweet}{rgb}{1.0, 0.44, 0.37}

\newcommand{\empirical}[1]{#1}

\newcommand{\point}[1]{{\noindent\bf #1:} }

\lstdefinelanguage{Rust}{%
  sensitive%
, morecomment=[l]{//}%
, morecomment=[s]{/*}{*/}%
, moredelim=[s][{\itshape\color[rgb]{0,0,0.75}}]{\#[}{]}%
, morestring=[b]{"}%
, alsodigit={}%
, alsoother={}%
, alsoletter={!}%
, morekeywords={break, continue, else, for, if, in, loop, match, return, while}  
, morekeywords={as, const, let, move, mut, ref, static}  
, morekeywords={dyn, enum, fn, impl, Self, self, struct, trait, type, union, use, where}  
, morekeywords={crate, extern, mod, pub, super}  
, morekeywords={unsafe}  
, morekeywords={abstract, alignof, become, box, do, final, macro, offsetof, override, priv, proc, pure, sizeof, typeof, unsized, virtual, yield}  
, morekeywords=[2]{Add, AddAssign, Any, AsciiExt, AsInner, AsInnerMut, AsMut, AsRawFd, AsRawHandle, AsRawSocket, AsRef, Binary, BitAnd, BitAndAssign, Bitor, BitOr, BitOrAssign, BitXor, BitXorAssign, Borrow, BorrowMut, Boxed, BoxPlace, BufRead, BuildHasher, CastInto, CharExt, Clone, CoerceUnsized, CommandExt, Copy, Debug, DecodableFloat, Default, Deref, DerefMut, DirBuilderExt, DirEntryExt, Display, Div, DivAssign, DoubleEndedIterator, DoubleEndedSearcher, Drop, EnvKey, Eq, Error, ExactSizeIterator, ExitStatusExt, Extend, FileExt, FileTypeExt, Float, Fn, FnBox, FnMut, FnOnce, Freeze, From, FromInner, FromIterator, FromRawFd, FromRawHandle, FromRawSocket, FromStr, FullOps, FusedIterator, Generator, Hash, Hasher, Index, IndexMut, InPlace, Int, Into, IntoCow, IntoInner, IntoIterator, IntoRawFd, IntoRawHandle, IntoRawSocket, IsMinusOne, IsZero, Iterator, JoinHandleExt, LargeInt, LowerExp, LowerHex, MetadataExt, Mul, MulAssign, Neg, Not, Octal, OpenOptionsExt, Ord, OsStrExt, OsStringExt, Packet, PartialEq, PartialOrd, Pattern, PermissionsExt, Place, Placer, Pointer, Product, Put, RangeArgument, RawFloat, Read, Rem, RemAssign, Seek, Shl, ShlAssign, Shr, ShrAssign, Sized, SliceConcatExt, SliceExt, SliceIndex, Stats, Step, StrExt, Sub, SubAssign, Sum, Sync, TDynBenchFn, Terminal, Termination, ToOwned, ToSocketAddrs, ToString, Try, TryFrom, TryInto, UnicodeStr, Unsize, UpperExp, UpperHex, WideInt, Write}
, morekeywords=[2]{Send}  
, morekeywords=[3]{bool, char, f32, f64, i8, i16, i32, i64, isize, str, u8, u16, u32, u64, unit, usize, i128, u128}  
, morekeywords=[4]{Err, false, None, Ok, Some, true}  
, morekeywords=[3]{AccessError, Adddf3, AddI128, AddoI128, AddoU128, ADDRESS, ADDRESS64, addrinfo, ADDRINFOA, AddrParseError, Addsf3, AddU128, advice, aiocb, Alignment, AllocErr, AnonPipe, Answer, Arc, Args, ArgsInnerDebug, ArgsOs, Argument, Arguments, ArgumentV1, Ashldi3, Ashlti3, Ashrdi3, Ashrti3, AssertParamIsClone, AssertParamIsCopy, AssertParamIsEq, AssertUnwindSafe, AtomicBool, AtomicPtr, Attr, auxtype, auxv, BackPlace, BacktraceContext, Barrier, BarrierWaitResult, Bencher, BenchMode, BenchSamples, BinaryHeap, BinaryHeapPlace, blkcnt, blkcnt64, blksize, BOOL, boolean, BOOLEAN, BoolTrie, BorrowError, BorrowMutError, Bound, Box, bpf, BTreeMap, BTreeSet, Bucket, BucketState, Buf, BufReader, BufWriter, Builder, BuildHasherDefault, BY, BYTE, Bytes, CannotReallocInPlace, cc, Cell, Chain, CHAR, CharIndices, CharPredicateSearcher, Chars, CharSearcher, CharsError, CharSliceSearcher, CharTryFromError, Child, ChildPipes, ChildStderr, ChildStdin, ChildStdio, ChildStdout, Chunks, ChunksMut, ciovec, clock, clockid, Cloned, cmsgcred, cmsghdr, CodePoint, Color, ColorConfig, Command, CommandEnv, Component, Components, CONDITION, condvar, Condvar, CONSOLE, CONTEXT, Count, Cow, cpu, CRITICAL, CStr, CString, CStringArray, Cursor, Cycle, CycleIter, daddr, DebugList, DebugMap, DebugSet, DebugStruct, DebugTuple, Decimal, Decoded, DecodeUtf16, DecodeUtf16Error, DecodeUtf8, DefaultEnvKey, DefaultHasher, dev, device, Difference, Digit32, DIR, DirBuilder, dircookie, dirent, dirent64, DirEntry, Discriminant, DISPATCHER, Display, Divdf3, Divdi3, Divmoddi4, Divmodsi4, Divsf3, Divsi3, Divti3, dl, Dl, Dlmalloc, Dns, DnsAnswer, DnsQuery, dqblk, Drain, DrainFilter, Dtor, Duration, DwarfReader, DWORD, DWORDLONG, DynamicLibrary, Edge, EHAction, EHContext, Elf32, Elf64, Empty, EmptyBucket, EncodeUtf16, EncodeWide, Entry, EntryPlace, Enumerate, Env, epoll, errno, Error, ErrorKind, EscapeDebug, EscapeDefault, EscapeUnicode, event, Event, eventrwflags, eventtype, ExactChunks, ExactChunksMut, EXCEPTION, Excess, ExchangeHeapSingleton, exit, exitcode, ExitStatus, Failure, fd, fdflags, fdsflags, fdstat, ff, fflags, File, FILE, FileAttr, filedelta, FileDesc, FilePermissions, filesize, filestat, FILETIME, filetype, FileType, Filter, FilterMap, Fixdfdi, Fixdfsi, Fixdfti, Fixsfdi, Fixsfsi, Fixsfti, Fixunsdfdi, Fixunsdfsi, Fixunsdfti, Fixunssfdi, Fixunssfsi, Fixunssfti, Flag, FlatMap, Floatdidf, FLOATING, Floatsidf, Floatsisf, Floattidf, Floattisf, Floatundidf, Floatunsidf, Floatunsisf, Floatuntidf, Floatuntisf, flock, ForceResult, FormatSpec, Formatted, Formatter, Fp, FpCategory, fpos, fpos64, fpreg, fpregset, FPUControlWord, Frame, FromBytesWithNulError, FromUtf16Error, FromUtf8Error, FrontPlace, fsblkcnt, fsfilcnt, fsflags, fsid, fstore, fsword, FullBucket, FullBucketMut, FullDecoded, Fuse, GapThenFull, GeneratorState, gid, glob, glob64, GlobalDlmalloc, greg, group, GROUP, Guard, GUID, Handle, HANDLE, Handler, HashMap, HashSet, Heap, HINSTANCE, HMODULE, hostent, HRESULT, id, idtype, if, ifaddrs, IMAGEHLP, Immut, in, in6, Incoming, Infallible, Initializer, ino, ino64, inode, input, InsertResult, Inspect, Instant, int16, int32, int64, int8, integer, IntermediateBox, Internal, Intersection, intmax, IntoInnerError, IntoIter, IntoStringError, intptr, InvalidSequence, iovec, ip, IpAddr, ipc, Ipv4Addr, ipv6, Ipv6Addr, Ipv6MulticastScope, Iter, IterMut, itimerspec, itimerval, jail, JoinHandle, JoinPathsError, KDHELP64, kevent, kevent64, key, Key, Keys, KV, l4, LARGE, lastlog, launchpad, Layout, Lazy, lconv, Leaf, LeafOrInternal, Lines, LinesAny, LineWriter, linger, linkcount, LinkedList, load, locale, LocalKey, LocalKeyState, Location, lock, LockResult, loff, LONG, lookup, lookupflags, LookupHost, LPBOOL, LPBY, LPBYTE, LPCSTR, LPCVOID, LPCWSTR, LPDWORD, LPFILETIME, LPHANDLE, LPOVERLAPPED, LPPROCESS, LPPROGRESS, LPSECURITY, LPSTARTUPINFO, LPSTR, LPVOID, LPWCH, LPWIN32, LPWSADATA, LPWSAPROTOCOL, LPWSTR, Lshrdi3, Lshrti3, lwpid, M128A, mach, major, Map, mcontext, Metadata, Metric, MetricMap, mflags, minor, mmsghdr, Moddi3, mode, Modsi3, Modti3, MonitorMsg, MOUNT, mprot, mq, mqd, msflags, msghdr, msginfo, msglen, msgqnum, msqid, Muldf3, Mulodi4, Mulosi4, Muloti4, Mulsf3, Multi3, Mut, Mutex, MutexGuard, MyCollection, n16, NamePadding, NativeLibBoilerplate, nfds, nl, nlink, NodeRef, NoneError, NonNull, NonZero, nthreads, NulError, OccupiedEntry, off, off64, oflags, Once, OnceState, OpenOptions, Option, Options, OptRes, Ordering, OsStr, OsString, Output, OVERLAPPED, Owned, Packet, PanicInfo, Param, ParseBoolError, ParseCharError, ParseError, ParseFloatError, ParseIntError, ParseResult, Part, passwd, Path, PathBuf, PCONDITION, PCONSOLE, Peekable, PeekMut, Permissions, PhantomData, pid, Pipes, PlaceBack, PlaceFront, PLARGE, PoisonError, pollfd, PopResult, port, Position, Powidf2, Powisf2, Prefix, PrefixComponent, PrintFormat, proc, Process, PROCESS, processentry, protoent, PSRWLOCK, pthread, ptr, ptrdiff, PVECTORED, Queue, radvisory, RandomState, Range, RangeFrom, RangeFull, RangeInclusive, RangeMut, RangeTo, RangeToInclusive, RawBucket, RawFd, RawHandle, RawPthread, RawSocket, RawTable, RawVec, Rc, ReadDir, Receiver, recv, RecvError, RecvTimeoutError, ReentrantMutex, ReentrantMutexGuard, Ref, RefCell, RefMut, REPARSE, Repeat, Result, Rev, Reverse, riflags, rights, rlim, rlim64, rlimit, rlimit64, roflags, Root, RSplit, RSplitMut, RSplitN, RSplitNMut, RUNTIME, rusage, RwLock, RWLock, RwLockReadGuard, RwLockWriteGuard, sa, SafeHash, Scan, sched, scope, sdflags, SearchResult, SearchStep, SECURITY, SeekFrom, segment, Select, SelectionResult, sem, sembuf, send, Sender, SendError, servent, sf, Shared, shmatt, shmid, ShortReader, ShouldPanic, Shutdown, siflags, sigaction, SigAction, sigevent, sighandler, siginfo, Sign, signal, signalfd, SignalToken, sigset, sigval, Sink, SipHasher, SipHasher13, SipHasher24, size, SIZE, Skip, SkipWhile, Slice, SmallBoolTrie, sockaddr, SOCKADDR, sockcred, Socket, SOCKET, SocketAddr, SocketAddrV4, SocketAddrV6, socklen, speed, Splice, Split, SplitMut, SplitN, SplitNMut, SplitPaths, SplitWhitespace, spwd, SRWLOCK, ssize, stack, STACKFRAME64, StartResult, STARTUPINFO, stat, Stat, stat64, statfs, statfs64, StaticKey, statvfs, StatVfs, statvfs64, Stderr, StderrLock, StderrTerminal, Stdin, StdinLock, Stdio, StdioPipes, Stdout, StdoutLock, StdoutTerminal, StepBy, String, StripPrefixError, StrSearcher, subclockflags, Subdf3, SubI128, SuboI128, SuboU128, subrwflags, subscription, Subsf3, SubU128, Summary, suseconds, SYMBOL, SYMBOLIC, SymmetricDifference, SyncSender, sysinfo, System, SystemTime, SystemTimeError, Take, TakeWhile, tcb, tcflag, TcpListener, TcpStream, TempDir, TermInfo, TerminfoTerminal, termios, termios2, TestDesc, TestDescAndFn, TestEvent, TestFn, TestName, TestOpts, TestResult, Thread, threadattr, threadentry, ThreadId, tid, time, time64, timespec, TimeSpec, timestamp, timeval, timeval32, timezone, tm, tms, ToLowercase, ToUppercase, TraitObject, TryFromIntError, TryFromSliceError, TryIter, TryLockError, TryLockResult, TryRecvError, TrySendError, TypeId, U64x2, ucontext, ucred, Udivdi3, Udivmoddi4, Udivmodsi4, Udivmodti4, Udivsi3, Udivti3, UdpSocket, uid, UINT, uint16, uint32, uint64, uint8, uintmax, uintptr, ulflags, ULONG, ULONGLONG, Umoddi3, Umodsi3, Umodti3, UnicodeVersion, Union, Unique, UnixDatagram, UnixListener, UnixStream, Unpacked, UnsafeCell, UNWIND, UpgradeResult, useconds, user, userdata, USHORT, Utf16Encoder, Utf8Error, Utf8Lossy, Utf8LossyChunk, Utf8LossyChunksIter, utimbuf, utmp, utmpx, utsname, uuid, VacantEntry, Values, ValuesMut, VarError, Variables, Vars, VarsOs, Vec, VecDeque, vm, Void, WaitTimeoutResult, WaitToken, wchar, WCHAR, Weak, whence, WIN32, WinConsole, Windows, WindowsEnvKey, winsize, WORD, Wrapping, wrlen, WSADATA, WSAPROTOCOL, WSAPROTOCOLCHAIN, Wtf8, Wtf8Buf, Wtf8CodePoints, xsw, xucred, Zip, zx}
, morekeywords=[5]{assert!, assert_eq!, assert_ne!, cfg!, column!, compile_error!, concat!, concat_idents!, debug_assert!, debug_assert_eq!, debug_assert_ne!, env!, eprint!, eprintln!, file!, format!, format_args!, include!, include_bytes!, include_str!, line!, module_path!, option_env!, panic!, print!, println!, select!, stringify!, thread_local!, try!, unimplemented!, unreachable!, vec!, write!, writeln!}  
}%

\newcommand{\tool}{{\sc Scythe}\xspace}
\newcommand{\perses}{{\sc Perses}\xspace}
\newcommand{\vulcan}{{\sc Vulcan}\xspace}
\newcommand{\latra}{{\sc Latra}\xspace}
\newcommand{\gp}{S$+$P\xspace}
\newcommand{\bcell}[1]{\cellcolor{gray!20}\textbf{#1}}

\definecolor{javaKw}{HTML}{0033B3}
\definecolor{javaCm}{HTML}{6F8DAB}
\definecolor{javaStr}{HTML}{067D17}
\definecolor{javaBg}{HTML}{FAFAFA}
\definecolor{javaRule}{HTML}{888888}
\lstdefinestyle{javapaper}{%
  language=Java,
  basicstyle=\scriptsize\fontfamily{lmtt}\selectfont,
  keywordstyle=\color{javaKw}\bfseries,
  commentstyle=\color{javaCm}\itshape,
  stringstyle=\color{javaStr},
  frame=leftline,            
  framerule=1.2pt,
  rulecolor=\color{javaKw!50},
  xleftmargin=10pt,
  numbers=none,
  showstringspaces=false,
  breaklines=true,
  columns=fullflexible,
  aboveskip=3pt,
  belowskip=3pt,
  literate={\{}{{\{}}1 {\}}{{\}}}1,
}

\lstdefinestyle{javapanel}{%
  language=Java,
  basicstyle=\footnotesize\fontfamily{lmtt}\selectfont,
  keywordstyle=\color{javaKw}\bfseries,
  commentstyle=\color{javaCm}\itshape,
  stringstyle=\color{javaStr},
  frame=leftline,
  framerule=0.8pt,
  rulecolor=\color{javaKw!50},
  xleftmargin=16pt,
  showstringspaces=false,
  breaklines=true,
  breakatwhitespace=true,
  postbreak=\mbox{\textcolor{javaRule}{$\hookrightarrow$}\space},
  columns=fullflexible,
  aboveskip=2pt,
  belowskip=2pt,
  literate={\{}{{\{}}1 {\}}{{\}}}1,
}

\usepackage{xspace}
\usepackage{tikz}
\usetikzlibrary{positioning,arrows.meta,shapes,fit}